\documentclass[conference]{IEEEtran}
\IEEEoverridecommandlockouts

\usepackage{cite}
\usepackage{amsmath,amssymb,amsfonts}
\usepackage{algorithmic}
\usepackage{graphicx}
\usepackage{textcomp}
\usepackage{xcolor}

\usepackage{acronym}
\usepackage{url}
\usepackage[hidelinks]{hyperref}
\usepackage{algorithm}
\usepackage{amssymb}
\usepackage{pifont}
\usepackage{booktabs}
\usepackage{enumitem}
\usepackage{amssymb}

\def\BibTeX{{\rm B\kern-.05em{\sc i\kern-.025em b}\kern-.08em
    T\kern-.1667em\lower.7ex\hbox{E}\kern-.125emX}}

\DeclareMathOperator*{\argmin}{arg\,min}
\newcommand{\cmark}{\ding{51}} 
\newcommand{\xmark}{\ding{55}} 

\begin{document}

\title{\textsc{QinCodec}: Neural Audio Compression with Implicit Neural Codebooks}

\author{\IEEEauthorblockN{Zineb Lahrichi}
\IEEEauthorblockA{\textit{Sony AI, Télécom Paris} \\
Paris, France}
\and
\IEEEauthorblockN{Gaëtan Hadjeres}
\IEEEauthorblockA{\textit{Sony AI} \\
Paris, France}
\and
\IEEEauthorblockN{Gaël Richard}
\IEEEauthorblockA{\textit{Télécom Paris} \\
Paris, France}
\and
\IEEEauthorblockN{Geoffroy Peeters}
\IEEEauthorblockA{\textit{Télécom Paris} \\
Paris, France}
}

\maketitle

\begin{abstract}
Neural audio codecs, neural networks which compress a waveform into discrete tokens, play a crucial role in the recent development of audio generative models. State-of-the-art codecs rely on the end-to-end training of an autoencoder and a quantization bottleneck.
However, this approach restricts the choice of the quantization methods as it requires to define how gradients propagate through the quantizer and how to update the quantization parameters online. In this work, we revisit the common practice of joint training and propose to quantize the latent representations of a pre-trained autoencoder offline, followed by an optional finetuning of the decoder to mitigate degradation from quantization. This strategy allows to consider any off-the-shelf quantizer, especially state-of-the-art trainable quantizers with implicit neural codebooks such as \textsc{Qinco2}. We demonstrate that with the latter, our proposed codec termed \textsc{QinCodec}, is competitive with baseline codecs while being notably simpler to train. Finally, our approach provides a general framework that amortizes the cost of autoencoder pretraining, and enables more flexible codec design.
\end{abstract}


\section{Introduction}

Traditional audio codecs, such as MP3 or Opus, are compression systems that encode digital audio into smaller bit-based intermediate representations, while enabling accurate reconstruction via a decoder.  In recent years, neural audio codecs have emerged as robust alternatives to traditional handcrafted approaches and are now essential to the creation of speech and audio autoregressive generative models \cite{borsos2023audiolmlanguagemodelingapproach,defossez2024moshi}.

 \begin{figure}[ht!]
\begin{minipage}[b]{1.0\linewidth}
  \centerline{\includegraphics[width=8cm]{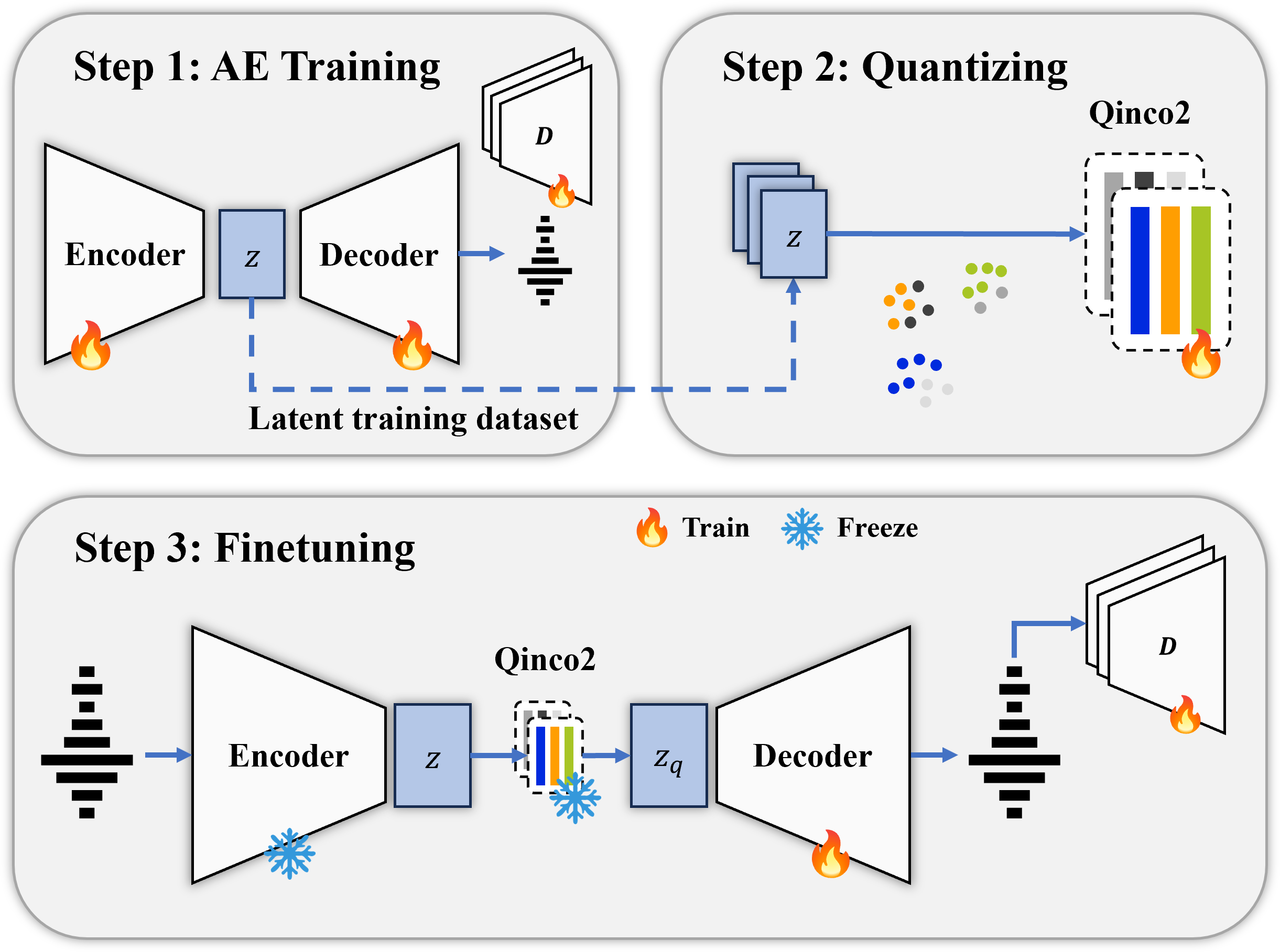}}
\end{minipage}
\caption{\textbf{Training procedure of \textsc{QinCodec} with offline quantization}: First, we train a continuous compression model with spectral and adversarial losses. Next, we quantize the bottleneck latent vectors into discrete embeddings. We then finetune the decoder on the quantized representations.
}
\label{fig:main figure}
\end{figure} 
 
Many advances in Neural Audio Codecs research are building upon the approach from \cite{zeghidour2021soundstreamendtoendneuralaudio} often denoted as RVQ-GAN, which consists in the end-to-end training of an autoencoder with a Residual Vector Quantization (RVQ) bottleneck layer.
 
However, one critical assumption within this framework has remained unchallenged: namely the need for \textit{end-to-end} training, which requires defining 1. how to propagate gradients (the quantization layer is non-differentiable by definition) and 2. how to perform online updates of the quantizer parameters, usually done via bespoke formulas.
These constraints significantly narrow the range of quantization layer designs since such operations may not always be possible.
In particular, \textsc{Qinco} \cite{huijben2024residual} and its follow-up \textsc{Qinco2} \cite{vallaeys2025qinco2vectorcompressionsearch} were proposed as powerful generalizations of unsupervised RVQ, relying on a trainable neural network that adapts codebooks implicitly for more accurate quantization. However, their complexity and specific training loop limit their suitability for online settings, as they are primarily designed for fixed dataset compression and retrieval.
In this paper, we propose a three-stages strategy that allows us to rely on \textsc{Qinco2} for neural audio coding:
\begin{itemize}
        \item We introduce \textsc{QinCodec}, a 44.1 kHz audio codec based on the decoupled training of an autoencoder and a neural residual vector quantizer \textsc{Qinco2}, trained offline.
        \item Our model is the first auto-encoder that relies on Vocos \cite{siuzdak2024vocosclosinggaptimedomain} blocks, providing a lightweight and fast way to encode/decode audio, making its integration easy into the training pipelines of generative models.
        \item \textsc{QinCodec} outperforms state-of-the-art methods at 16 kbps bitrate and achieves competitive results at 8kbps.
        with both objective and subjective metrics.
        \item Our offline approach offers a simple yet robust framework that allows to consider any off-the-shelf quantizer with a fixed pre-trained autoencoder, paving the way for adaptable and frugal codec design.  
\end{itemize}
We believe that this work is the first to demonstrate the viability of non-end-to-end training for audio codecs and hope that this may drive attention to research in offline quantization methods and their applications for generative modeling. 
A website with audio examples is available at \url{https://zinebl-sony.github.io/post-training-rvq/}.

\section{Related Work}
\label{sec:related-work}

\subsection{Vector Quantization techniques}

Vector Quantization (VQ) is the task of encoding continuous data into a discrete set of vector codes and is fundamental to data compression \cite{DBLP:journals/corr/abs-1711-00937} and approximate nearest neighbor search \cite{6619223}. Traditional VQ employs classical clustering algorithms, but its computational cost scales linearly with the number of codes, prompting the need for low-distortion alternatives. Multi-codebook approaches, such as Residual Vector Quantization (RVQ) \cite{DBLP:journals/corr/MartinezHL14}, alleviate this by progressively refining quantization from coarse to fine. Lastly, neural quantizers have leveraged deep networks to compress data more efficiently—targeting large-scale database compression and rapid retrieval. Notably, Unsupervised Neural Quantization \cite{morozov2019unsupervisedneuralquantizationcompresseddomain} combines deep networks with Product Quantization and the Gumbel-softmax trick, while \textsc{Qinco} \cite{huijben2024residualquantizationimplicitneural} adapts residual codes within the RVQ framework.

\subsection{Neural Audio Codecs} 

Recent advances in neural audio coding take inspiration from earlier approaches for learning discrete representations using deep neural autoencoders. A prominent class of these methods leverages continuous noisy relaxations of discrete variables such as Concrete relaxations \cite{maddison2017concretedistributioncontinuousrelaxation} using the Gumbel-Softmax distribution \cite{jang2017categoricalreparameterizationgumbelsoftmax}, enabling optimization via backpropagation. 
Alternatively, Softmax quantization was also considered for speech coding \cite{DBLP:journals/corr/abs-1710-09064}. 

These methods were progressively replaced by the widely adopted Vector Quantization Variational Autoencoders (VQ-VAEs). This class of models utilizes a straight-through estimator to pass gradients from the decoder to the encoder \cite{DBLP:journals/corr/abs-1711-00937}, 
together with Exponential Moving Average (EMA) updates of the codes based on their assigned vectors. In the audio compression domain, \cite{DBLP:journals/corr/abs-2107-03312} introduced RVQ-GANs, by integrating RVQ in the VAE-GAN framework \cite{larsen2016autoencodingpixelsusinglearned}. This hierarchical quantization approach enables raw waveform compression at variable bitrates, while ensuring high-quality synthesis through the combined use of adversarial, reconstruction, and codebook losses.
Multiple approaches \cite{kumar2023highfidelityaudiocompressionimproved,défossez2022highfidelityneuralaudio} improved upon this RVQ-GAN framework by addressing some of its major limitations such as low codebook usage, loss balancing and difficult hyperparameter tuning. This can be tackled, for instance, with architectural improvements \cite{ahn2024hilcodechighfidelitylightweight}, alternative quantization methods \cite{mentzer2023finitescalarquantizationvqvae}, 
propagating gradients with a variant of the straight-through estimator \cite{fifty2024restructuring}, and more sophisticated discriminators \cite{bak2023avocodogenerativeadversarialnetwork}. 

Now, the most recent development in audio compression involves scaling encoder-decoder architectures by incorporating transformer blocks before and after the quantization layer while retaining RVQ, as seen in models like the Mimi codec \cite{defossez2024moshi} or by using Finite Scalar Quantization (FSQ) \cite{parker2024scaling}. 

\section{Residual Vector Quantization (RVQ)}
\label{sec: RVQ}

\subsection{Conventional RVQ}
\label{sec: rvq}

Vector Quantization refers to the mapping of continuous data embeddings to discrete vectors selected from a finite set, called a \textit{codebook}, with its size specified in bits. To this end, clustering heuristics like k-means \cite{arthur2006k} are used, but these methods exhibit poor scalability when applied to large codebooks. RVQ overcomes this limitation by using a sequence of smaller codebooks, where the residual quantization error is iteratively quantized by the next codebook. As in \cite{huijben2024residual}, we set some notations to formalize RVQ and its variants throughout the paper. We aim to quantize the vectors $x \in \mathbb{R}^D$ by using a sequence of codebooks \( (\bar{C}_1, \dots, \bar{C}_N) \), each containing $K$ entries $( \bar{c}_i^1, \dots, \bar{c}_i^K )_{i\in \{1,N\}}$. 
Let $\hat{x}_n$ be the reconstruction after $n-1$ steps, with $ \hat{x}_1 = 0$. Quantization proceeds iteratively, where at each step $n$, the residual vector $r_n = x - \hat{x}_n $ is encoded by selecting the closest entry from $\bar{C}_n$. 

\subsection{\textsc{Qinco}: Implicit Neural Codebooks}
\label{sec:Qinco}

A fundamental limitation of RVQ, is its use of static codebooks at each quantization stage, ignoring residual error distributions shaped by earlier codebooks. To address that, they introduce a neural residual vector quantizer that depends on prior intermediate reconstructions, and where the codebooks are learned \textit{implicitely} via a neural network.

At each quantization step $n$, a neural network $f_{\theta_{n}}$ generates a specialized codebook centroid $c_n^k = f_{\theta_{n}}(\hat{x}_n, \bar{c}_n^{k})$, conditioned on the previous reconstruction $\hat{x}_n$ and an initial centroid $\bar{c}_n^{k}$ taken from a pre-trained base codebook $\bar{C}_{n}$ obtained with k-means clustering as described in~\ref{sec: rvq}. 

For each centroid, an affine transformation projects the concatenation of $\bar{c}_n^k$ and $\hat{x}_n$, followed by $L$ residual blocks. In addition, they set $f_{\theta_1}$ to identity to compensate the null conditioning $\hat{x}_1$ at the initial step, resulting in $C_1 = \bar{C}_1$. Residual connections enable base codebooks $(\bar{C}_{1}...\bar{C}_{n})$ to propagate through the network, allowing it to achieve and outperform conventional RVQ baselines.

The conditioning procedure forces the decoding to be sequential with the update rule $\hat{x}_{n+1} \leftarrow \hat{x}_n + f_{\theta_n}(\hat{x}_n, \bar{c}_n^{k})$. 
During training the parameters are learned via stochastic gradient descent to minimize the sum of mean squared errors between the residuals and centroids over all centroids and codebooks.

\subsection{Improved Residual Vector Quantization (iRVQ)}
\label{sec:irvq}
 
We propose a straightforward variant of RVQ, by adding residual conditioning as in \textsc{Qinco}, relying on residual statistics and re-standardization. This quantization, which we introduce for our ablation study, can be seen as a special case of \textsc{Qinco} without neural network training. More precisely, at each step $n$, the residual $r_{n+1}$ after quantization of $r_n$ is
\begin{equation}
r_{n+1} = (r_n - c)/\sigma_c \quad \textrm{with} \quad c := \argmin_{\bar{c}^k_n \in  \bar{C}_n, k \in \{1...K\}}||r_n - \bar{c}^k_n||^2_2
\end{equation}
where $\sigma_{c} \in \mathbb{R}^{D}$ is the per-cluster standard deviation of all inputs assigned to cluster $c$. The final approximation $\hat{x}$ can be reconstructed with the update rule $\hat{x}_{n+1} \leftarrow \hat{x}_n + \sigma_c c$,
without re-centering to ensure decreasing quantization errors with respect to the number of codebooks.
It should be noted that such standardization procedure may be complex to implement in end-to-end autoencoder training as one would have to update these statistics online.

\textbf{Null codebook} \label{ss:null} Since RVQ is additive, we fix a null vector to each codebook (except the first one) to ensure the decrease of the quantization error after each quantization step in RVQ and avoid the addition of random noise. 

\section{\textsc{QinCodec}}
\label{sec:aposteriorirvq}

\subsection{Step 1: Autoencoder pre-training}
\label{ae architecture}

Our proposed autoencoder borrows ideas from Vocos \cite{siuzdak2024vocosclosinggaptimedomain}, a GAN-based vocoder, trained to produce STFT coefficients from an audio signal. The decoder is a succession of ConvNeXt blocks followed by an iSTFT head and our encoder consists in its mirrored architecture namely a complex STFT layer followed by ConvNeXt blocks. Finally, a linear bottleneck is inserted between the encoder and the decoder. The primary advantage of this architecture is its consistent temporal resolution across all layers, avoiding artifacts from up-sampling \cite{DBLP:journals/corr/abs-2010-14356}. Additionally, since convolutions are performed on sequences of uniform length $ \mathrm{sample \ size} / \mathrm{hop \ length}$, the architecture ensures fast training.

The discriminators we use are identical to the ones used in \cite{kumar2023highfidelityaudiocompressionimproved} and our objective includes a spectral loss, an adversarial loss and a feature matching loss. More details are provided below in  \ref{training details}

\begin{table*}[t]
   \centering
    \caption{Quantitative evaluation of the proposed model at 16 kbps compared to competing baselines.}
    \begin{tabular}{{lccccccccc}}                
        \toprule
         & Quantizer & Finetuning &\textbf{Si-SDR} $\uparrow$ & \textbf{MS-Mel} $\downarrow$ & \textbf{FD-OL3} $\downarrow$ & \textbf{FM-OL3} $\downarrow$ & \textbf{Perplexity} $\uparrow$ \\
        \hline
        \textsc{DAC}                &   &                   & \textbf{9.04} & 0.85 & 51.4 & 0.23 & 784 \\
        \textsc{EnCodec}            &   &                   & 6.10 & 1.39 & 97.2 & 0.35 & 588 \\
        \hline
                            & iRVQ &  \xmark                 & 6.20 & 0.93 & 45.5 & 0.23 & 926 \\
                            & iRVQ &  \cmark                 & 6.58 & 0.82 & 31.3 & 0.21 & 926 \\
                            & \textsc{Qinco2} & \xmark                & 7.22 & 0.79 & 38.1 & 0.21 & 980 \\
        \textsc{QinCodec}   & \textsc{Qinco2} & \cmark                & 7.55 & \textbf{0.74} & \textbf{34.4} & \textbf{0.19} & \textbf{980} \\
        \bottomrule
    \end{tabular}
    \label{tab: baselines_16}
\end{table*}

\begin{table*}[t]
    \centering
    \caption{Quantitative evaluation of the proposed model at 8 kbps compared to competing baselines.}
    \begin{tabular}{{lccccccccc}}
        \toprule
         & Quantizer & Finetuning &\textbf{Si-SDR} $\uparrow$ & \textbf{MS-Mel} $\downarrow$ & \textbf{FD-OL3} $\downarrow$ & \textbf{FM-OL3} $\downarrow$ & \textbf{Perplexity} $\uparrow$ \\
        \hline
        \textsc{DAC}                &   &                   & \textbf{5.49} & \textbf{1.07} & 45.7 & 0.28 & 746 \\
        \textsc{EnCodec}            &   &                   & 3.22 & 1.57 & 98.4 & 0.39 & 506 \\
        \hline
                            & iRVQ &  \xmark                 & 2.60 & 1.46 & 77.0 & 0.33 & 919 \\
                            & iRVQ &  \cmark                 & 3.77 & 1.14 & 39.3 & 0.28 & 919 \\
                            & \textsc{Qinco2} & \xmark                & 3.96 & 1.32 & 67.1 & 0.31 & 957 \\
        \textsc{QinCodec}   & \textsc{Qinco2} & \cmark                & 4.64 & 1.12 & \textbf{35.7} & \textbf{0.27} & \textbf{957} \\
        \bottomrule
    \end{tabular}
    \label{tab: baselines_8}
\end{table*}

\subsection{Step 2: Offline Vector Quantization}
\label{sec: offline vq}

After training the compression model, we leverage its latent representations to train a residual vector quantizer in an offline setting. We consider a set of audio embeddings $\mathcal{Z}$ deriving from our encoder $\mathcal{E}$ and audio inputs. Each embedding lies in $\mathbb{R}^{D \times T}$, where $T$ and $D$ 
are respectively the number of frames and the latent dimension of our model. From $\mathcal{Z}$, we form a collection $\mathcal{X}$ of vector frames in $\mathbb{R}^D$, which are subsequently used to train the residual vector quantizer. 
In this case, the number of quantizers $N$ determines the target $ \mathrm{bitrate} = N \times \log_2{K} \times F $, where $ F = T / d$
is the latent frame rate, $d$ the duration of the input audio and $\log_2{K}$ is the size of each codebook in binary bits.
The offline vector quantization step is generally lightweight and only takes a fraction of the time of the pre-training or finetuning stages.

\subsection{Step 3: Finetuning}
\label{sec:finetuning}

In the finetuning stage, we freeze the encoder and the quantizer and back-propagate only through the decoder and the discriminators. In other words, we improve the previously trained GAN-vocoder in accomplishing the task of decoding the quantized representation. 

\section{Experimental setup}
\label{sec:experiments}

\subsection{Training details}
\label{training details}
For the first step of our framework, we train three compression models with latent dimension $D = \{16, 32, 64\}$, with the architecture described in~\ref{ae architecture}. Our three models are trained using the AdamW optimizer with weight decay of $\mathrm{1e^{-3}}$ and a learning rate of $\mathrm{2e^{-4}}$ including exponential warmup and decay. Betas are set to $\mathrm{(0.5, 0.9)}$. Each model is trained for 1M steps across 8 A100 GPUs, with an effective batch size of $\mathrm{240}$.

Our inputs are complex STFTs, configured with $\mathrm{n_{fft} = 512}$, a hop length of $\mathrm{256}$ and input audio duration is $d=1$s.
As in the original Vocos model, we use 8 ConvNeXt layers for both the encoder and the decoder. For the discriminator, we combine the complex multi-scale STFT and the Multi-Period Discriminator (MPD), as in \cite{kumar2023highfidelityaudiocompressionimproved}. 
The training objective includes a spectral reconstruction $L_1$ loss based on mel-coefficients as in \cite{kumar2023highfidelityaudiocompressionimproved}, an adversarial loss using the LSGAN formulation \cite{DBLP:journals/corr/MaoLXLW16}, and a feature matching loss as formulated in \cite{kumar2023highfidelityaudiocompressionimproved}. The corresponding weights for these losses are set to $\mathrm{15}$, $\mathrm{1}$, and $\mathrm{2}$, respectively. 

For the finetuning step, we keep the same configuration as for training, except for the learning rate which is aligned with the learning rate of the last training epoch.

\subsection{Baselines}
\label{sec:baselines}
We evaluate the performance of \textsc{Qincodec} against two baselines: the pretrained 44.1kHz \textsc{DAC}
and 48kHz \textsc{EnCodec} models at 8kbps and 16kbps. Both are RVQ-GANs, following the same framework as \cite{zeghidour2021soundstreamendtoendneuralaudio} and trained on general sounds among other types like speech and music. However, they differ in their codebook update strategies, loss-balancing techniques, and architectural designs. 

\subsection{Datasets}

We train and validate our models on 1-second, 44.1KHz audio clips from WavCaps \cite{wavcaps}, a dataset of 400k general sounds from AudioSet, FreeSound, BBC Sound Effects, and SoundBible. For quantization, we randomly sample audios from the training set, extract their latent representations, and shuffle them. We use around 5M frames for conventional RVQ and 60M frames for training \textsc{Qinco2}. Evaluation is conducted with 5-second audio chunks from the AudioCaps \cite{audiocaps} test set, ensuring no overlap with the training data and consistency with the baselines evaluation protocols.

\subsection{Objective and subjective metrics}

We evaluate reconstruction quality using objective and perceptual metrics. For objective metrics, we measure the Scale-Invariant Signal-to-Distortion Ratio (Si-SDR) and multi-scale mel reconstruction error, as defined in \ref{training details}. For perceptual metrics, we compute the Fréchet Distance and feature matching with OpenL3 \cite{8682475}.
To assess quantization quality in the bottleneck, we measure the Perplexity i.e the entropy of the codebook's representation, which reflects how uniformly codebook entries are used. 

We conducted a MUSHRA test \footnote{\url{https://github.com/audiolabs/webMUSHRA}}
to compare the perceptual quality of \textsc{QinCodec} with \textsc{DAC} at 8 and 16 kbps. The test included 12 randomly selected 5-second excerpts spanning various sound categories, with a hidden reference and a 3.5 kHz low-pass filtered anchor. 

\subsection{Quantization details}
\noindent
\textbf{RVQ and iRVQ}: We utilize the mini batched implementation of k-means\footnote{\url{https://scikit-learn.org/stable/modules/generated/sklearn.cluster.MiniBatchKMeans.html}} to scale to a few million vectors. We adapted the batch size to three times the size of the codebooks and otherwise keep the default parameters.

\noindent
\textbf{\textsc{Qinco2}}: The original paper provides detailed guidance on how to select hyperparameters that optimize various quantizer attributes, including precision (low MSE), encoding/decoding efficiency, and training speed. For our experiments, we prioritize a balanced tradeoff between training efficiency and reconstruction accuracy, selecting a medium-sized model with $L=4$ residual blocks and a hidden dimension of $d_e = d_h = 384$. Regarding optimization, we use the default parameters specified in the original paper.

\begin{figure}[t]
    \centering
  \includegraphics[width=0.6\linewidth]{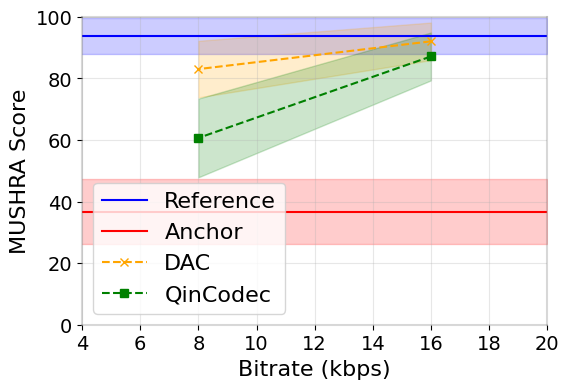}
  \caption{MUSHRA scores with 95\% confidence intervals for \textsc{DAC}and \textsc{QinCodec} and a 3.5kHz low-pass anchor, evaluated at 8 kbps and 16 kbps.}
    \label{fig:mushra}
\end{figure} 

\section{Results and discussions}

\subsection{Comparison with baselines} 
\label{sec: comparison}

\begin{figure}[b]
    \centering
     \includegraphics[width=0.6\linewidth]{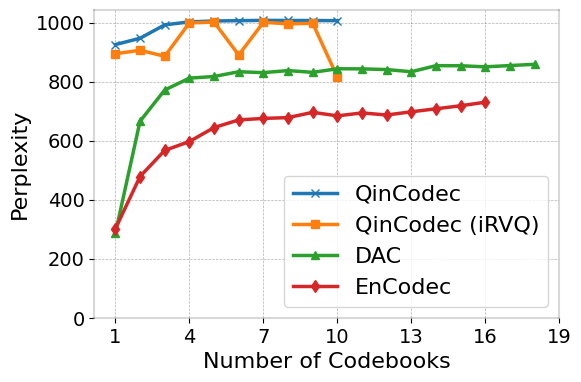}
     \caption{Perplexity vs. number of codebooks for \textsc{EnCodec}, \textsc{DAC}, \textsc{QinCodec}, and \textsc{QinCodec} with iRVQ quantization.}
    \label{fig:perplexity}
\end{figure}

\textsc{QinCodec} outperforms \textsc{DAC} and \textsc{EnCodec} at 16 kbps across all metrics except Si-SDR, where \textsc{DAC} scores higher (see \autoref{tab: baselines_16} and Fig.\ref{fig:mushra}), likely due to the phase information loss in the quantization layer. Fig.\ref{fig:perplexity} shows that \textsc{QinCodec} achieves higher and more consistent perplexity scores across codebooks, highlighting the benefit of offline-trained quantizers for codebook optimization. At 8 kbps, the gap between \textsc{DAC} and \textsc{QinCodec} narrows, leading to similar performance (see \autoref{tab: baselines_8}).

MUSHRA scores align with SI-SDR, with \textsc{DAC} slightly ahead but within overlapping confidence intervals. Both models surpass the 3.5 kHz low-pass anchor. While \textsc{QinCodec} matches or exceeds baselines in objective metrics, its perceived quality remains limited by the quantizer capacity, particularly at lower bitrates.

\subsection{Ablation studies}

\autoref{tab: baselines_16} shows the perceptual improvements gained from finetuning after offline quantization. A few extra epochs of finetuning enhance perceptual metrics like FD-OL3, with minimal impact on spectral metrics, highlighting the importance of finetuning, even briefly, to optimize perceptual quality.

We also evaluate our model with different offline quantizers: iRVQ and \textsc{Qinco2} (see \autoref{tab: latentMSE}). iRVQ improves code precision over RVQ, likely by handling outliers through re-standardization. \textsc{Qinco2} outperforms both, with gains reflected in audio metrics at 8 and 16 kbps, demonstrating that a more expressive codebook enhances reconstruction quality.

\subsection{Influence of the latent dimension}

\begin{figure}[t]
\centering
        \centering
        \includegraphics[width=0.8\linewidth]{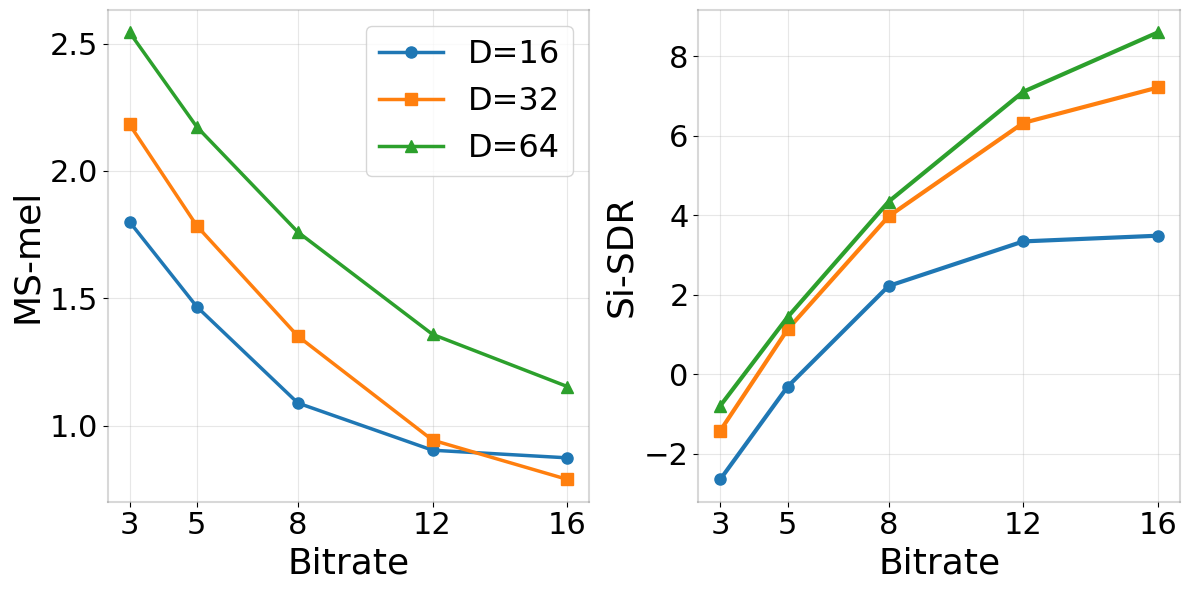} 
\caption{Performance of \textsc{QinCodec} at various bitrates, and latent dimensions.}
\label{fig:latentdimension}
\end{figure}

Increasing the latent dimension $D$ improves reconstruction fidelity by embedding more information in continuous models, but these gains do not directly translate to offline quantization, which is inherently limited by the bitrate. Fig.~\ref{fig:latentdimension} shows that higher $D$ increases MS-mel  (since quantization becomes more complex and lossy), but also improves SI-SDR by encoding richer embeddings, including phase information. In conclusion, the latent dimension $D$ represents the trade-off between fidelity and compression rate and has to be chosen carefully (In our case, $D=32$ seems to be the optimal choice).

\begin{table}[t]
\centering
\caption{Performance of offline quantizers at 16 kbps, after step 2.}
\begin{tabular}{lccc}
    \hline
        & \textbf{MSE (z)} $\downarrow$ & \textbf{Si-SDR} $\uparrow$ & \textbf{MS-Mel} $\downarrow$\\
        \hline
        RVQ             & 2.29 &  6.09  & 0.96 \\
        iRVQ            & 2.23 &  6.20  & 0.93 \\
        \textsc{Qinco2} & \textbf{1.51} & \textbf{7.22} & \textbf{0.79} \\
        \hline
            \end{tabular}
\label{tab: latentMSE}
\end{table}

\section{CONCLUSION}

We show that end-to-end training is not a prerequisite for vector-quantized neural audio compression. Our codec, \textsc{QinCodec}, leverages a novel three-step training procedure with offline quantization to eliminate complex gradient propagation while enhancing quantization performance. This approach allows to use any off-the-shelf quantizer like \textsc{Qinco} without optimization constraints, yielding competitive results against end-to-end baselines. However, a trade-off between distortion, compression rate, and latent dimension limits performance at lower bitrates. In future work, we will narrow the gap between end-to-end and modular approaches and develop offline quantizers optimized for audio coding, capitalizing on the training stability of our method to scale Neural Audio Coding architectures.

\bibliographystyle{IEEEtran}
\bibliography{IEEEabrv, main}

\end{document}